\documentclass[reprint,
amsmath,amssymb,
 aps,
prb
floatfix,
]{revtex4-2}
\usepackage{graphicx,amsmath}
\usepackage{hyperref,capt-of,subcaption}
\usepackage{dcolumn,braket}
\usepackage{bm}
\usepackage{xcolor}
\usepackage{amsfonts}
\usepackage[utf8]{inputenc}

\begin{document}

\preprint{APS/123-QED}

\title{Single magnon excited states of a Heisenberg spin-chain using a quantum computer}

\author{Shashank Kumar Ranu}
\affiliation{Department of Electrical Engineering, Indian Institute of Technology Madras, Chennai, India}
	\affiliation{Department of Physics, Indian Institute of Technology Madras, Chennai, India}
\email{ee16s300@ee.iitm.ac.in}	
\author{Daniel D. Stancil}%
\affiliation{Department of Electrical and Computer Engineering, NC State University Raleigh, USA}
 \email{ddstancil@ncsu.edu}

\date{\today}

\begin{abstract}
Excited states of spin-chains play an important role in condensed matter physics. 
We present a method of calculating the single magnon excited states of the Heisenberg spin-chain that can be efficiently implemented on a quantum processor for small spin chains.
Our method involves finding the stationary points of the energy vs wavenumber curve. We implement our method for 4-site and 8-site Heisenberg Hamiltonians using numerical techniques as well as using an IBM quantum processor. Finally, we give an insight into the circuit complexity and scaling of our proposed method.
\end{abstract}
\maketitle
\section{\label{sec:Intro}Introduction}
We are currently in the era of Noisy Intermediate Scale Quantum (NISQ) devices wherein limited coherence time, gate imperfections, and readout errors affect the performance of quantum computers \cite{preskill2018quantum}. Though these NISQ devices are not 
fault-tolerant, they have been used to show quantum advantage over classical computers  \cite{arute2019quantum,zhong2020quantum} for some calculations. 
However, we are still far away from achieving Feynman's vision of simulating a large and complex quantum system using a quantum computer. Hybrid classical-quantum algorithms have been proposed to use these NISQ devices for practical applications. For example, Variational Quantum Eigensolver (VQE), a hybrid classical-quantum algorithm, is used to find the approximate ground state of a system with Hamiltonian $H$ \cite{peruzzo2014variational}. Shallow circuit depth and resistance to noise make VQE an attractive algorithm in this NISQ era. However, the shallow circuit depth of VQE comes at the cost of an increase in the repetitions of the circuit. Furthermore, VQE requires a classical optimization loop which makes the overall computation cumbersome.

Ground state energy plays a vital role in condensed matter physics and quantum chemistry, thereby leading to intense research on VQE and its variants in recent years. However, apart from the ground state, excited states of a system also play an important role in chemical reactions and physical processes. Various approaches such as Subspace Search Variational Quantum Eigensolver (SS-VQE), Von-Neumann entropy method (``WAVES"), and the variational quantum deflation method have been proposed to study the excited states of the system \cite{SSVQE,higgott2019variational,Santagati_2018}. However, these excited state estimation methods require deep circuits as they have VQE or quantum phase estimation as a sub-routine. They also require a classical optimization loop. Some of these methods such as SS-VQE and weighted SS-VQE require more measurements compared to a typical VQE implementation, further adding to the computational cost.

Knowledge of excited states also helps in calculating two-time correlation functions. These time correlation functions can be used to study the dynamics of spin waves. Recent works use an extension of the Hadamard test to calculate the correlation function of the form $\langle\hat{A}(t)\hat{B}(0)\rangle$ for the Heisenberg spin-chain \cite{francis2020quantum}. This calculation involves the preparation of the ground state of the Heisenberg system followed by the application of a controlled-$\hat{B}$ operator. An ancilla in a uniform superposition state acts as a control for the controlled-$\hat{B}$ operation. This step splits the wavefunction into ground and excited states which subsequently undergo time evolution as per the Heisenberg Hamiltonian. The final step involves a controlled-$\hat{A}$ operation and measurement of the ancilla qubit in a suitable basis to obtain the real and imaginary part of the correlation function. Note that this controlled-$\hat{A}$ operation between the ancilla and the Heisenberg site of interest typically involves a series of SWAP gates due to the limited connectivity of NISQ processors. Furthermore, the time evolution of the Heisenberg Hamiltonian
may require as many as
three CNOTs for each pair-wise interaction term of the Hamiltonian. Lack of all-to-all connectivity further necessitates the use of SWAP gates in the time evolution circuit. The presence of a large number of SWAP and CNOT gates makes scaling of this method a challenging problem. Post-processing of the raw data is needed to get meaningful results even for a 4-site Heisenberg spin-spin chain. 

Our work aims to provide a computationally efficient way of estimating the excited state of a Heisenberg spin chain. We use a parameterized circuit ($U(\phi)$) to prepare the trial first excited state of the Heisenberg Hamiltonian. We obtain the expectation of the Heisenberg Hamiltonian for the trial wavefunction for different $\phi$ values. Unlike VQE, our method does not involve a classical optimization feedback loop. We obtain the excited state by finding the saddle points of the energy expectation values Vs $\phi$ curve. This work also has a reduced circuit complexity compared to \cite{francis2020quantum} as it does not require the time evolution operator and Hadamard tests to estimate the single magnon excited states of the Heisenberg chain. 
\section{\label{sec:Prelims}Preliminaries}
\subsection{\label{subsec:trial wvfn} Single magnon wavefunction}
We focus on a linear ferromagnetic chain with $n$ sites. The Heisenberg Hamiltonian for a 1-D chain of spins is
\begin{equation}
H=-\frac{J}{2}\sum_i\boldsymbol{\sigma}_i\cdot\boldsymbol{\sigma}_{i+1},\label{eq:Heisenberg ham}
\end{equation}
where $J$ is the exchange coupling constant and $\boldsymbol{\sigma}_i$ is the Pauli vector representing the quantum spin operator at site $i$. The ground state of a linear ferromagnetic chain has all spins aligned parallel. 
Intuitively, we can write the lowest excited states as a superposition of states having a single flipped spin:
\begin{equation}
    \ket{\psi}_{N-\text{site}}=\sum_{n=0}^{N-1} C_n \ket{n},\label{eq:1st excited state}
\end{equation}
where $\ket{n}$ is the basis state with flipped state at the $n^{\text{th}}$ site and $\vert C_n \vert^2$ denotes the probability of finding the $N$-site ferromagnetic chain in the state $\ket{n}$. As spin waves are delocalized over all the sites, the probability of observing a flipped spin at any site is equal to $\frac{1}{N}$.  Furthermore, we assume that the probability amplitudes differ only in phase, but with progressive phase shift between the sites owing to the wave nature of the excitation. Hence, we write the probability amplitude as
\begin{equation}
    C_n=\frac{1}{\sqrt{N}}e^{in\phi} \label{eq:co-efficients}.
\end{equation}
For example, we write 
a single-flipped-spin (single magnon)
excited state of the $4$-site ferromagnetic Heisenberg chain using Eq.~\eqref{eq:1st excited state} and Eq.~\eqref{eq:co-efficients} as
\begin{eqnarray}
\ket{\psi}_{4-\text{site}}&=&\frac{1}{2}\big[\ket{0001}+e^{i\phi}\ket{0010}+e^{i2\phi}\ket{0100}\nonumber\\
&&\;\;\;\;+e^{i3\phi}\ket{1000}\big].\label{eq:4-site trial state}
\end{eqnarray}
\subsection{Stationary points of the energy Vs $\phi$ curve }\label{subsec:proof}
Let $\ket{\psi_n}$ be an energy eigenstate of the ferromaganetic Heisenberg Hamiltonian. We assume that our method gives an approximate eigenstate of the system such that
\begin{equation}
    \ket{\tilde{\psi}_n}=\ket{\psi_n}+\ket{\delta_n},
\end{equation}
where $\ket{\delta_n}$ captures a small deviation from the $n^{\text{th}}$ correct solution. Hence, an estimate of the energy of the $n^\text{th}$ eigenstate is
\begin{eqnarray}
\bra{\tilde{\psi}_n}H\ket{\tilde{\psi}_n}&=&\left(\bra{\psi_n}+\bra{\delta_n}\right)H\left(\ket{\psi_n}+\ket{\delta_n}\right)\nonumber \\
&=&\bra{\psi_n}H\ket{\psi_n}+\bra{\psi_n}H\ket{\delta_n}+\bra{\delta_n}H\ket{\psi_n}\nonumber \\
&&+\bra{\delta_n}H\ket{\delta_n}.\label{eq:exp}
\end{eqnarray}
Here $\bra{\psi_n}H\ket{\psi_n}$ is the correct excited state energy $E_n$. 
Since our ansatz for the eigenstates (Eq.~\eqref{eq:1st excited state} and Eq.~\eqref{eq:co-efficients}) contains a single adjustable parameter $\phi$ (equivalent to the normalized wave number), we now show that the terms linear in $\phi$ vanish as $\phi\rightarrow\phi_n$, so that the curve of estimated energy (Eq.~\eqref{eq:exp}) vs $\phi$ is stationary around the correct eigenstates.

Since $\bra{\delta_n}H\ket{\psi_n}$ and $\bra{\psi_n}H\ket{\delta_n}$ are complex conjugates of one another, it is sufficient to show that the terms linear in $\phi-\phi_n$ from these quantities are imaginary as $\phi\rightarrow\phi_n$.
We now show that this is the case for our trial wave function.

Based on our ansatz, the correct expression for the $n^\text{th}$ eigenstate can be written
\begin{equation}
    \ket{\psi_n}=\frac{1}{\sqrt{N}}\sum_{m=0}^{N-1} e^{im\phi_n}\ket{m},\label{eq:correct}
\end{equation}
where $\phi_n$ is the correct value of $\phi$ for the $n^\text{th}$ eigenstate. Our approximation to the eigenstate can therefore be written
\begin{equation}
    \ket{\tilde{\psi}_n}=\frac{1}{\sqrt{N}}\sum_{m=0}^{N-1} e^{im(\phi_n+\Delta)}\ket{m}=\ket{\psi_n}+\ket{\delta_n},\label{eq:approximate}
\end{equation}
where $|\Delta|\ll 1$. Solving for $\ket{\delta_n}$ gives
\begin{equation}
    \ket{\delta_n}= \frac{1}{\sqrt{N}}\sum_{m=0}^{N-1} e^{im\phi_n}\left[e^{im\Delta}-1\right]\ket{m}.\label{eq:error term}
\end{equation}
For $|\Delta|\ll 1$ this simplifies to
\begin{equation}
    \ket{\delta_n}\approx \frac{i\Delta}{\sqrt{N}}\sum_{m=0}^{N-1} me^{im\phi_n}\ket{m}.\label{eq:linear error term}
\end{equation}

To evaluate the dependence of the error terms in Eq.~\eqref{eq:exp} on $\Delta$, it will be convenient to consider the slightly more general term $\bra{\delta_n}H\ket{\psi_k}$. We have
\begin{eqnarray}
    \bra{\delta_n}H\ket{\psi_k} &=& E_k\braket{\delta_n|\psi_k}\nonumber\\
    &=& -E_k\frac{i\Delta}{N} \sum_{m,m'}m\,e^{i(m'\phi_k-m\phi_n)}\braket{m|m'}\nonumber\\
    &=& -E_k\frac{i\Delta}{N}\sum_{m=0}^{N-1}m\,e^{im(\phi_k-\phi_n)}.\label{eq:linear delta term}
\end{eqnarray}
For the specific case $k=n$ this reduces to
\begin{eqnarray}
    \bra{\delta_n}H\ket{\psi_n} 
    &=& -E_n\frac{i\Delta}{N}\sum_{m=0}^{N-1}m\nonumber\\
    &=&-iE_n\Delta\frac{N-1}{2}.\label{eq:linear delta term 2}
\end{eqnarray}
We see that to lowest order this term is indeed imaginary, so that there is no contribution to the energy estimate that is linear in $\Delta$ from these terms.

We next turn to the expression $\bra{\delta_n}H\ket{\delta_n}$ to determine its lowest-order dependence on $\Delta$. Using the property that $\sum_k\ket{\psi_k}\bra{\psi_k}=1$, we can write
\begin{eqnarray}
 \bra{\delta_n}H\ket{\delta_n} &=& \sum_{k=0}^{N-1}\bra{\delta_n}H\ket{\psi_k}\braket{\psi_k|\delta_n}\nonumber\\
 &=&\sum_{k=0}^{N-1}E_k\left|\braket{\delta_n|\psi_k}\right|^2.\label{eq:quadratic delta term 1}
\end{eqnarray}
Noting that the expression for $\braket{\delta_n|\psi_k}$ can be obtained from Eq.~\eqref{eq:linear delta term}, we have finally
\begin{equation}
    \bra{\delta_n}H\ket{\delta_n} = \frac{\Delta^2}{N^2}\sum_{k=0}^{N-1}E_k\left|\sum_{m=0}^{N-1}m\,e^{im(\phi_k-\phi_n)}\right|^2.
\end{equation}
Consequently we see that there is no first-order term in $\Delta$ from this term either, and we conclude that the energy vs $\phi$ curve must be stationary around the values corresponding to the eigenstates of the system.

Note that we have not consistently found all of the terms that are second-order in $\Delta$; we have only shown that the error terms vanish as $\Delta\rightarrow 0$. As described in the next section, numerically we find that the second derivative is positive around eigenstate values when $\phi_n<\pi/2$, negative around eigenstate values when $\phi_n>\pi/2$, and zero when $\phi_n=\pi/2$. Consequently we will find that eigenstates correspond to minima when $\phi_n<\pi/2$, maxima when $\phi_n>\pi/2$, and a saddle point when $\phi_n=\pi/2$.
\section{Implementation}
\subsection{Numerical implementation}
\begin{figure}[h!]
				\centerline{\includegraphics[width=0.35\textwidth,height=\textheight,keepaspectratio]{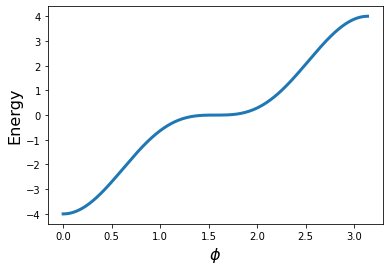}}
				\caption{Energy Vs $\phi$ curve for 4-site ferromagnetic Heisenberg Hamiltonian. In this and all following energy plots, the energy is normalized to $J/2$.}
				\label{fig:4-site energy}
			\end{figure}

\begin{figure}[h!]
				\centerline{\includegraphics[width=0.35\textwidth,height=\textheight,keepaspectratio]{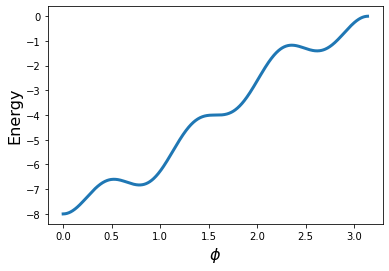}}
				\caption{Energy Vs $\phi$ curve for 8-site ferromagnetic Heisenberg Hamiltonian}
				\label{fig:8-site energy}
			\end{figure}

We numerically obtain the expected energy for the 
single-magnon excited states
of the ferromagnetic Heisenberg system using the NumPy module of Python. First, we expand the terms in the Heisenberg Hamiltonian (see Eq.~\eqref{eq:Heisenberg ham}) for a 4-site Heisenberg spin chain. The expanded Hamiltonian is
\begin{eqnarray}
H &=& -\frac{J}{2}(\sigma_0^x\otimes\sigma_1^x+\sigma_0^y\otimes\sigma_1^y+\sigma_0^z\otimes\sigma_1^z\nonumber \\
&&\;\;\;+\sigma_1^x\otimes\sigma_2^x +\sigma_1^y\otimes\sigma_2^y+\sigma_1^z\otimes\sigma_2^z\nonumber\\
&&\;\;\;+\sigma_2^x\otimes\sigma_3^x +\sigma_2^y\otimes\sigma_3^y+\sigma_2^z\otimes\sigma_3^z\nonumber\\
&&\;\;\;+\sigma_3^x\otimes\sigma_0^x +\sigma_3^y\otimes\sigma_0^y+\sigma_3^z\otimes\sigma_0^z).\label{eq:expanded ham}
\end{eqnarray}
Note that the term $\sigma_3^x\otimes\sigma_0^x$ is really a shorthand for $\sigma_3^x\otimes I \otimes I\otimes \sigma_0^x$, where we are taking the least significant qubit (rightmost) as $0$ and the most significant qubit (leftmost) as $3$. We can write this in shorthand as
\begin{equation}
\sigma_3^x\otimes I \otimes I\otimes \sigma_0^x=XIIX.
\end{equation}
In terms of this notation, the Hamiltonian shown in Eq.~\eqref{eq:expanded ham} can be written
\begin{eqnarray}
H &=&-\frac{J}{2}( IIXX + IIYY+ IIZZ\nonumber\\
&&\;\;\;+ IXXI + IYYI + IZZI\nonumber\\
&&\;\;\;+XXII + YYII + ZZII\nonumber\\
&&\;\;\;+ XIIX + YIIY + ZIIZ).\label{eq:shorthand ham}
\end{eqnarray}
We expand the terms in the Heisenberg Hamiltonian for an 8-site Heisenberg spin chain in a similar manner. As shown in Eq.~\eqref{eq:co-efficients}, our single-magnon excited state wavefunction is parametric in nature. So, we vary $\phi$ from $0$ to $\pi$ in steps of $\frac{\pi}{800}$ and obtain a plot of expected energy Vs $\phi$. Fig.~\ref{fig:4-site energy} and Fig.~\ref{fig:8-site energy} show the energy Vs $\phi$ curve for the 4-site and 8-site ferromagnetic Heisenberg spin chain. 

\begin{figure}[h!]
				\centerline{\includegraphics[width=0.35\textwidth,height=\textheight,keepaspectratio]{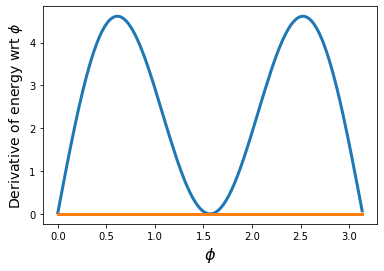}}
				\caption{Derivative of energy with respect to $\phi$ vs $\phi$ curve for 4-site ferromagnetic Heisenberg Hamiltonian}
				\label{fig:4-site energy derivative}
			\end{figure}

\begin{figure}[h!]
				\centerline{\includegraphics[width=0.35\textwidth,height=\textheight,keepaspectratio]{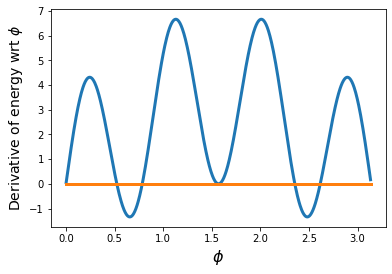}}
				\caption{Derivative of energy with respect to $\phi$ vs $\phi$ curve for 8-site ferromagnetic Heisenberg Hamiltonian}
				\label{fig:8-site energy derivative}
			\end{figure}

As discussed in Subsection~\ref{subsec:proof}, we are interested in the stationary points of the energy Vs $\phi$ curve. So, we obtain the derivative of expected energy with respect to $\phi$ using the forward finite difference method, and search for zeros. Fig~\ref{fig:4-site energy derivative} and Fig.~\ref{fig:8-site energy derivative} show the derivative of energy Vs $\phi$ for  the 4-site and 8-site ferromagnetic Heisenberg spin chain. We observe from Fig~\ref{fig:4-site energy derivative} and Fig.~\ref{fig:8-site energy derivative} that the energy derivative becomes zero for the 4-site ferromagnetic spin chain at $\phi=0,1.570,3.14$, while the derivative vanishes at $\phi=0$, $0.5216$, $0.7834$, $1.570$, $2.3542$, $2.616$ and $3.14$ for the 8-site spin chain. The values of $\phi$ at which the derivative of the energy Vs $\phi$ curve becomes $0$ are the potential solutions. 
Note that we have shown that the curve is stationary about the energies of valid excited states, 
but we have not shown that each stationary point corresponds to an excited state.
To test candidate points for validity, we must check to ensure that the boundary conditions of the ferromagnetic spin chain are satisfied.
The presence of terms of the form $\sigma_3^x\otimes\sigma_0^x$, $\sigma_3^y\otimes\sigma_0^y$ and $\sigma_3^z\otimes\sigma_0^z$ in Eq.~\eqref{eq:expanded ham} shows that a coupling exists between the first and the last site of our linear spin chain, thereby making the boundary condition periodic. As discussed in Subsection~\ref{subsec:trial wvfn}, the probability amplitude $C_n$ for finding the $N$-site ferromagnetic chain in the state $\ket{n}$ is proportional to $e^{i n\phi}$. Hence, for an $N$-site ferromagnetic Heisenberg Hamiltonian, the periodic boundary condition is satisfied when $C_0$ equals $C_{N}$, i.e., when
\begin{eqnarray}
 e^{iN\phi}&=&1,\nonumber\\
 \text{or}\quad \phi_n &=&\frac{2\pi n}{N},\quad n=0,1,2...N-1.
 \label{eq:bc}
\end{eqnarray}
From Fig.~\ref{fig:4-site energy derivative}, we find that when $\phi=0$, $1.570$ and $3.14$, the derivative of energy with respect to $\phi$ vanishes for the 4-site spin chain. These values are in agreement with Eqn. \ref{eq:bc} which gives $\phi=0$, $\pi/2$ and $\pi$.
(Note that the value for $n=3$ is $3\pi/2$. This is an equivalent point to $\phi=-\pi/2$ and simply represents a spin wave traveling in the opposite direction.)

\begin{figure}[h!]
\centering
\begin{subfigure}{0.2\textwidth}
  \includegraphics[width=\linewidth]{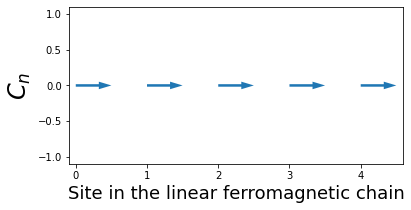}
  \caption{}
  \label{fig:1st maxima}
\end{subfigure}\hfil 
\begin{subfigure}{0.2\textwidth}
  \includegraphics[width=\linewidth]{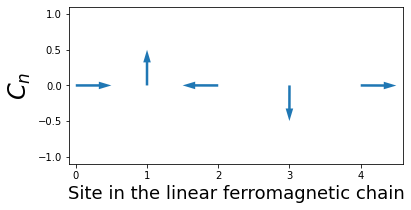}
  \caption{}
  \label{fig:ist minima}
\end{subfigure}\hfil 
\begin{subfigure}{0.2\textwidth}
  \includegraphics[width=\linewidth]{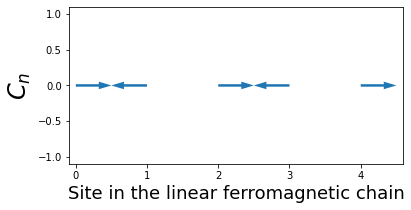}
  \caption{}
  \label{fig:saddle point}
\end{subfigure}\hfil 
\caption{Boundary condition for the extrema and saddle point of 4-site ferromagnetic Heisenberg Hamiltonian. a) $\{C_n\}$ for $\phi=0$ b) $\{C_n\}$ for $\phi=1.57$ c) $\{C_n\}$ for $\phi=3.14$.}
\label{fig:4-site BC}
\end{figure}

To illustrate satisfaction of the boundary conditions, we plot $\{C_n\}$ for these three values of $\phi$ in Fig.~\ref{fig:4-site BC}. The horizontal dimension of each vector shown in the polar plot (see Fig.~\ref{fig:4-site BC}) represents the real part of $\{C_n\}$, and the vertical dimension captures the imaginary part of the probability amplitude. We observe that the wave function satisfies the boundary condition (i.e., the values for site 0 equal those for site 4) for $\phi=0$, $1.570$ and $3.14$ (see Fig.~\ref{fig:4-site BC}) for the 4-site spin chain.

Similarly, the derivative of energy with respect to $\phi$ becomes $0$ when $\phi=0$, $0.5216$, $0.7834$, $1.570$, $2.3542$, $2.616$ and $3.14$ for the 8-site chain. 
From Eqn. \ref{eq:bc} the values satisfying the boundary conditions are $\phi=0$, $\pi/4$, $\pi$, $3\pi/4$, and $\pi$. Consequently the values of $0.5216$ and $2.616$ correspond to spurious solutions.

\begin{figure}[h!]
\centering
\begin{subfigure}{0.2\textwidth}
  \includegraphics[width=\linewidth]{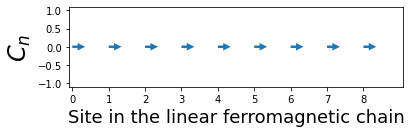}
  \caption{}
  \label{fig:1st maxima}
\end{subfigure}\hfil
\begin{subfigure}{0.2\textwidth}
  \includegraphics[width=\linewidth]{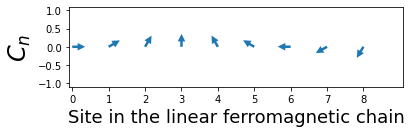}
  \caption{}
  \label{fig:1st maxima}
\end{subfigure}\hfil 
\begin{subfigure}{0.2\textwidth}
  \includegraphics[width=\linewidth]{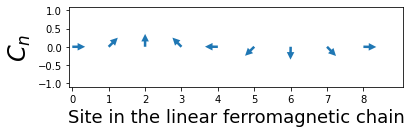}
  \caption{}
  \label{fig:ist minima}
\end{subfigure}\hfil 
\begin{subfigure}{0.2\textwidth}
  \includegraphics[width=\linewidth]{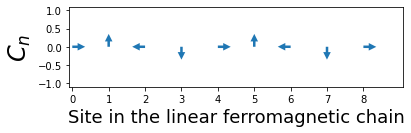}
  \caption{}
  \label{fig:saddle point}
\end{subfigure}\hfil 
\begin{subfigure}{0.2\textwidth}
  \includegraphics[width=\linewidth]{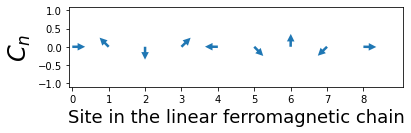}
  \caption{}
  \label{fig:2nd maxima}
\end{subfigure}\hfil 
\begin{subfigure}{0.2\textwidth}
  \includegraphics[width=\linewidth]{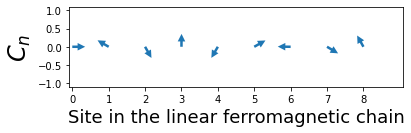}
  \caption{}
  \label{fig:2nd minima}
\end{subfigure}
\begin{subfigure}{0.2\textwidth}
  \includegraphics[width=\linewidth]{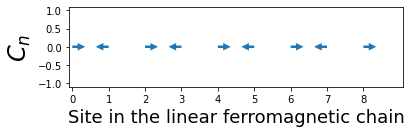}
  \caption{}
  \label{fig:1st maxima}
\end{subfigure}\hfil
\caption{Boundary condition for the stationary points of 8-site ferromagnetic Heisenberg Hamiltonian. a) $\{C_n\}$ for $\phi=0$ b) $\{C_n\}$ for $\phi=0.5216$ c) $\{C_n\}$ for $\phi=0.7834$ d) $\{C_n\}$ for $\phi=1.570$ e) $\{C_n\}$ for $\phi=2.3542$ f) $\{C_n\}$ for $\phi=2.616$ g) $\{C_n\}$ for $\phi=3.14$.}
\label{fig:BC 8-site}
\end{figure}

We show the polar plot of $\{C_n\}$ in Fig.~\ref{fig:BC 8-site} for the 7 different values of $\phi$ where the derivative is $0$. We observe that $\phi = 0.5216$ and $2.616$ do not satisfy the boundary condition, as expected. 
\subsection{Ansatz construction}
Calculating the single-magnon excited states using an IBM quantum computer involves preparing the parametrized state of Eq.~\eqref{eq:4-site trial state} using quantum gates. We use the approach based on Schmidt decomposition to construct the circuit needed to prepare the single-magnon state wavefunction \cite{2011}. We briefly outline this Schmidt decomposition based approach of preparing any arbitrary quantum state in this subsection.

We assume that the number of qubits under consideration ($n$) is even, and define $k=\frac{n}{2}$. We can represent a pure state $\ket{\psi}\in \mathcal{H}^n$ in the computational basis as
\begin{equation}
    \ket{\psi}=\sum_{i,j=1}^{2^k} C_{ij}\ket{a_i}\otimes\ket{b_j},\label{eq:matrix_form}
\end{equation}
where $\ket{a_i}\in \mathcal{H}^k$ and $\ket{b_j}\in \mathcal{H}^k$. Singular value decomposition (SVD) of the coefficient matrix $C$ gives
\begin{equation}
    C=U\Sigma V^{\dagger},\label{eq:general SVD}
\end{equation}
where $\Sigma$ is a diagonal matrix. Schmidt decomposition divides the Hilbert space of $n$ qubits into two parts, each containing $k$ qubits. Hence, Schmidt decomposition allows us to write an arbitrary pure state $\ket{\psi}$ as
\begin{equation}
    \ket{\psi}_{\text{SD}}=\sum_{i}^{2^k} \lambda_{i}\ket{\alpha_i}\otimes\ket{\beta_i},\label{eq:general SD}
\end{equation}
where $\lambda_{i}$ are the Schmidt coefficients, and $\ket{\alpha_i}$ and $\ket{\beta_i}$ form an orthogonal basis (Schmidt basis) in their respective Hilbert space. From Eq.~\eqref{eq:matrix_form}, Eq.~\eqref{eq:general SVD} and Eq.~\eqref{eq:general SD} one can show that $\lambda_{i}$ are the diagonal elements of $\Sigma$, $\alpha_i$ are the columns of $U$ and $\beta_i$ are the rows of $V^{\dagger}$.
As qubits in the IBM processors are initialized to the $\ket{0}$ state, preparing an arbitrary state $\ket{\psi}$ reduces to transforming $\ket{0}^{\otimes n}$ to $\ket{\psi}_{\text{SD}}$.
\subsubsection{Ansatz for constructing single-magnon excited state wave functions for the $4$-site Heisenberg Hamiltonian}\label{subsubsec:ansatz 4-site}
\begin{figure}[h!]
				\centerline{\includegraphics[width=0.45\textwidth,height=\textheight,keepaspectratio]{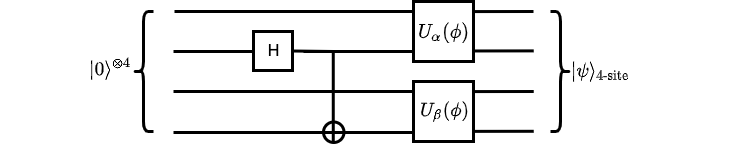}}
				\caption{Circuit for mapping $\ket{0}^{\otimes 4}$ to the  first excited state of $4$-site ferromagnetic Heisenberg chain.}
				\label{fig:4-site circuit}
			\end{figure}

We now show how to transform $\ket{0}^{4}$ to the 4-site single-magnon excited state ansatz (see Eq.~\eqref{eq:4-site trial state}). First, we obtain the coefficient matrix $C$ of the target state as shown below
\begin{equation}
    C_{\text{4-site}}=\frac{1}{2}\begin{pmatrix}
0 & 1 & e^{i\phi} & 0\\
e^{i2\phi} & 0 & 0 & 0\\
e^{i3\phi} & 0 & 0 & 0\\
0 & 0 & 0 & 0\\
\end{pmatrix}.\label{eq: Coeff_4-site}
\end{equation}
The SVD of $C_{\text{4-site}}$ gives
\begin{eqnarray}
&&U=\begin{pmatrix}
0 & e^{i\phi} & 0 & 0\\
\frac{e^{i2\phi}}{\sqrt{2}} & 0 & 0 & -\frac{e^{-i\phi}}{\sqrt{2}}\\
\frac{e^{i3\phi}}{\sqrt{2}} & 0 & 0 & \frac{1}{\sqrt{2}}\\
0 & 0 & 1 & 0\\
\end{pmatrix},\nonumber \\
&&\Sigma=\begin{pmatrix}
\frac{1}{\sqrt{2}} & 0 & 0 & 0\\
0 & \frac{1}{\sqrt{2}} & 0 & 0\\
0 & 0 & 0 & 0\\
0 & 0 & 0 & 0\\
\end{pmatrix},\nonumber \\
&&V=\begin{pmatrix}
1 & 0 & 0 & 0\\
0 & \frac{e^{i\phi}}{\sqrt{2}} & 0 & -\frac{e^{i\phi}}{\sqrt{2}}\\
0 & \frac{1}{\sqrt{2}} & 0 & \frac{1}{\sqrt{2}}\\
0 & 0 & 1 & 0\\
\end{pmatrix}.\label{eq:SVD 4-site}
\end{eqnarray}
The first step towards transforming $\ket{0}^{\otimes n}$ to an arbitrary state ($\ket{\psi}$) requires us to construct a circuit that does the mapping
\begin{equation}
    \ket{0}^{\otimes n}\to \sum_{i=1}^{2^k}\left(\lambda_i\ket{a_i}\right)\otimes\ket{0}^{\otimes k},\label{eq:4-site step 1a}
\end{equation}
where $\ket{a_i}\in \mathcal{H}^k$ forms the computational basis and $\lambda_i$ are the Schmidt coefficients which are equal to the diagonal entries of the $\Sigma$. Hence, for generating $\ket{\psi}_{\text{4-site}}$, the mapping as shown in Eq.~\eqref{eq:4-site step 1a} can be easily achieved by applying $I\otimes H\otimes I\otimes I$ 
\begin{equation}
    \ket{0000}\xrightarrow{I\otimes H\otimes I\otimes I} \frac{1}{\sqrt{2}}\left(\ket{00}+\ket{01}\right)\otimes\ket{00}.\label{eq4-site step 1b}
\end{equation}
The second step in the process of obtaining $\ket{\psi}$ from $\ket{0}^{\otimes n}$ consists of a transformation as shown below
\begin{equation}
 \sum_{i=1}^{2^k}\left(\lambda_i\ket{a_i}\right)\otimes\ket{0}^{\otimes k}\to \sum_{i=1}^{2^k}\lambda_i\ket{a_i}\otimes\ket{a_i}.\label{eq:4-site step 2a}
 \end{equation}
 A single CNOT gate between qubit $2$ and $4$ implements the required transformation of Eq.~\eqref{eq:4-site step 2a} in the case of $\ket{\psi}_{\text{4-site}}$.
 \begin{align}
  \frac{1}{\sqrt{2}}\left(\ket{00}+\ket{01}\right)\otimes\ket{00}&\xrightarrow{CNOT_{2\to 4}}\nonumber\\ &\frac{1}{\sqrt{2}}(\ket{00}\otimes\ket{00}
  +\ket{01}\otimes\ket{01}),\label{eq:4-site step 2b}
 \end{align}
where $CNOT_{2\to 4}$ represents a CNOT gate with qubit $2$ as control qubit and qubit $4$ as target qubit. The next step of this Schmidt decomposition based approach is to construct a circuit that transforms the computational basis ($\ket{a_i}$) of the first $k$ qubits to the Schmidt basis $\ket{\alpha}_i$, i.e,
\begin{equation}
 \sum_{i=1}^{2^k}\lambda_i\ket{a_i}\otimes\ket{a_i}\to  \sum_{i=1}^{2^k}\lambda_i\ket{\alpha_i}\otimes\ket{a_i}.\label{eq:4-site step 3a} 
\end{equation}
A unitary matrix that does the transformation of Eq.~\eqref{eq:4-site step 3a} when $\ket{\psi}_{\text{4-site}}$ is the target state is
\begin{equation}
U_\alpha(\phi)=\begin{pmatrix}
0 & e^{i\phi} & 0 & 0\\
\frac{e^{i2\phi}}{\sqrt{2}} & 0 & 0 & -\frac{e^{-i\phi}}{\sqrt{2}}\\
\frac{e^{i3\phi}}{\sqrt{2}} & 0 & 0 & \frac{1}{\sqrt{2}}\\
0 & 0 & 1 & 0\\
\end{pmatrix}.\label{eq:4-site step 3b} 
\end{equation}
$U_\alpha(\phi)$ acts on the state shown in Eq.~\eqref{eq:4-site step 2b} as shown below
\begin{eqnarray}
 \frac{1}{\sqrt{2}}(\ket{00}&&\otimes\ket{00}+\ket{01}\otimes\ket{01})\xrightarrow{U_\alpha(\phi)}\frac{1}{\sqrt{2}}\Big[\Big(\frac{e^{i2\phi}}{\sqrt{2}}\ket{01}\nonumber\\
 &&+\frac{e^{i3\phi}}{\sqrt{2}}\ket{10}\Big)\otimes\ket{00}+\frac{e^{i\phi}}{\sqrt{2}}\ket{00}\otimes\ket{01}\Big].\label{eq:4-site step 3c}
\end{eqnarray}
The final step of the process is to construct a circuit that maps the computational basis ($\ket{a_i}$) of the last $k$ qubits to the Schmidt basis $\ket{\beta}_i$, i.e,
\begin{equation}
\sum_{i=1}^{2^k}\lambda_i\ket{\alpha_i}\otimes\ket{a_i}\to \sum_{i=1}^{2^k}\lambda_i\ket{\alpha_i}\otimes\ket{\beta_i}. \label{eq:4-site step 4a}
\end{equation}
We obtain the unitary matrix that carries out the transformation of Eq.~\eqref{eq:4-site step 4a} when $\ket{\psi}_{\text{4-site}}$ is the target state as
\begin{equation}
U_\beta(\phi)=\begin{pmatrix}
1 & 0 & 0 & 0\\
0 & \frac{e^{-i\phi}}{\sqrt{2}} & 0 & -\frac{e^{-i\phi}}{\sqrt{2}}\\
0 & \frac{1}{\sqrt{2}} & 0 & \frac{1}{\sqrt{2}}\\
0 & 0 & 1 & 0\\
\end{pmatrix}\label{eq:4-site step 4b} 
\end{equation}
The action of $U_\beta(\phi)$ on the state shown in Eq.~\eqref{eq:4-site step 3c} is
\begin{eqnarray}
 &&\frac{1}{\sqrt{2}}\Big[\Big(\frac{e^{i2\phi}}{\sqrt{2}}\ket{01}+\frac{e^{i3\phi}}{\sqrt{2}}\ket{10}\Big)\otimes\ket{00}\frac{e^{i\phi}}{\sqrt{2}}\ket{00}\nonumber\\
 &&\otimes\ket{01}\Big]\xrightarrow{U_\beta(\phi)}\frac{1}{2}\big[\ket{0001}+e^{i\phi}\ket{0010}+e^{i2\phi}\ket{0100}\nonumber\\
 &&\;\;\;\;\;\;\;\;\;\;\;\;\;\;\;\;\;\;\;\;\;\;\;\;\;\;\;\;\;\;+e^{i3\phi}\ket{1000}\big].\label{eq:4-site step 4c}
\end{eqnarray}
We observe from Eq.~\eqref{eq:4-site step 4c} that the final state of this Schmidt decomposition based process gives the required first excited state parametric function. Fig.~\ref{fig:4-site circuit} shows the ansatz for the $4$-site ferromagnetic Heisenberg chain. 
\subsubsection{Ansatz for constructing single-magnon excited state wave functions for the $8$-site Heisenberg Hamiltonian}\label{subsubsec:ansatz 8-site}
We can write the first excited state of $8$-site ferromagnetic Heisenberg chain using Eq.~\eqref{eq:1st excited state} and Eq.~\eqref{eq:co-efficients} as
\begin{eqnarray}
\ket{\psi}_{8-\text{site}}&=&\frac{1}{2\sqrt{2}}\big[\ket{00000001}+e^{i\phi}\ket{00000010}+e^{i2\phi}\times\nonumber\\
&&\;\;\;\ket{00000100}+e^{i3\phi}\ket{00001000}+e^{i4\phi}\times\nonumber\\
&&\;\;\;\ket{00010000}+e^{i5\phi}\ket{00100000}+e^{i6\phi}\times\nonumber\\
&&\;\;\;\ket{01000000}+e^{i7\phi}\ket{10000000}\big].\label{eq:excited_8-site}
\end{eqnarray}
We again use the Schmidt decomposition based procedure described in the previous subsection to transform $\ket{0}^{\otimes 8}$ to the single-flipped-spin excited state ansatz for an $8$-site ferromagnetic Heisenberg chain. The details of the calculation are shown in Appendix~\ref{appendix:ansatz 8-site}. Fig.~\ref{fig:8-site circuit} shows the ansatz for the $8$-site ferromagnetic Heisenberg chain.
\begin{center}
\begin{figure}[h!]
				\centerline{\includegraphics[width=0.45\textwidth,height=\textheight,keepaspectratio]{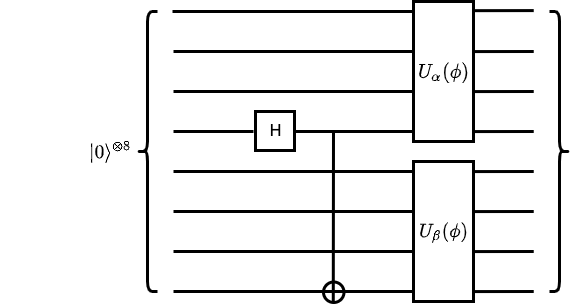}}
				\caption{Circuit for mapping $\ket{0}^{\otimes 8}$ to the  first excited state of $8$-site ferromagnetic Heisenberg chain.}
				\label{fig:8-site circuit}
			\end{figure}
\end{center}
\subsection{Implementation on IBM quantum computer}
\begin{center}
\begin{figure}[htbp!]
				\centerline{\includegraphics[width=0.45\textwidth,height=\textheight,keepaspectratio]{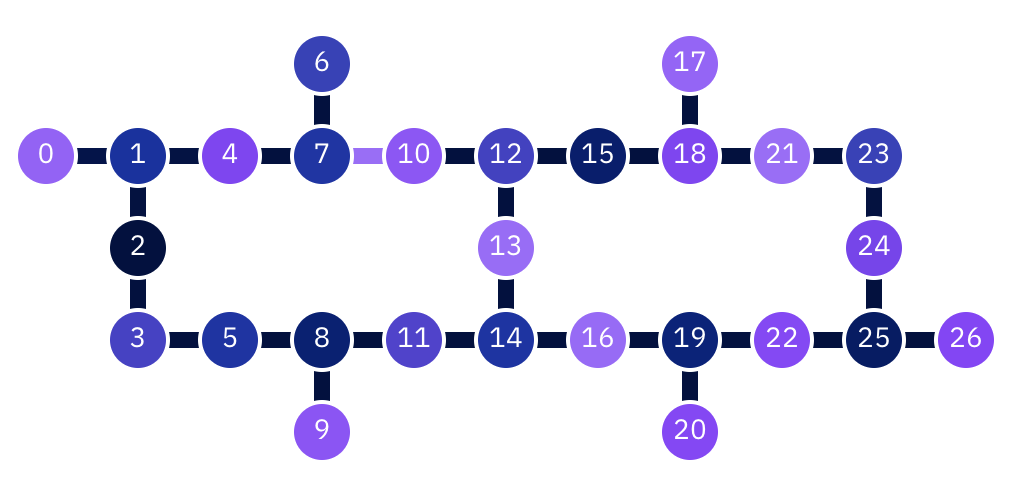}}
				\caption{Qubit connectivity in IBM Mumbai quantum processor \cite{backend}}
				\label{fig:connectivity}
			\end{figure}
\end{center}

\begin{figure}[h!]
    \centering
    \begin{subfigure}[t]{0.2\textwidth}
        \centering
        \includegraphics[width=\linewidth]{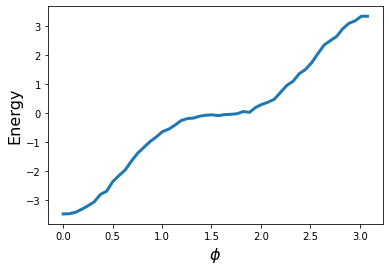}
        \caption{}\label{fig:4-site w/o correction}
    \end{subfigure}%
    ~ 
    \begin{subfigure}[t]{0.2\textwidth}
        \centering
        \includegraphics[width=\linewidth]{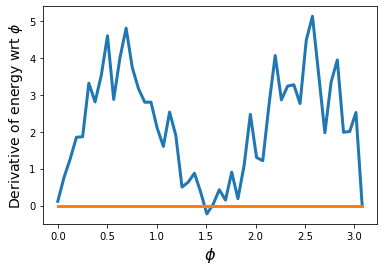}
        \caption{}\label{fig:grad_wo_correction}
    \end{subfigure}
    \caption{Results of 4-site ansatz implementation with noisy data points (a) Energy vs $\phi$ curve  (b) $\frac{\partial{H}}{\partial{\phi}}$ Vs $\phi$ curve.}\label{fig:wo_correction}
    \end{figure}

\begin{figure}[h!]
    \centering
    \begin{subfigure}[t]{0.2\textwidth}
        \centering
        \includegraphics[width=\linewidth]{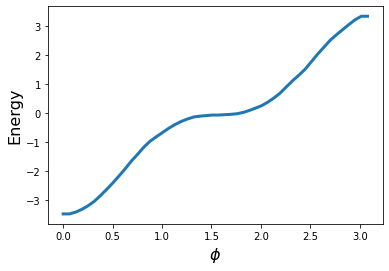}
        \caption{}\label{fig:4-site with correction}
    \end{subfigure}%
    ~ 
    \begin{subfigure}[t]{0.2\textwidth}
        \centering
        \includegraphics[width=\linewidth]{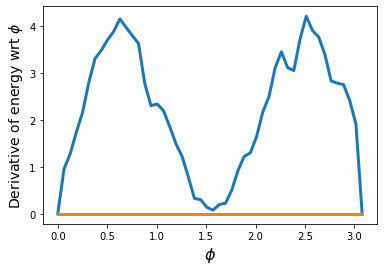}
        \caption{}\label{fig:grad_with_correction}
    \end{subfigure}
    \caption{Results of 4-site ansatz implementation with filtered data points (a) Energy vs $\phi$ curve  (b) $\frac{\partial{H}}{\partial{\phi}}$ Vs $\phi$ curve.}\label{fig:with_correction}
    \end{figure}

We used IBM Mumbai, a 27 qubit quantum processor with a quantum volume of 128 to implement our ansatz \cite{backend}. Fig.~\ref{fig:connectivity} shows the architecture of the IBM Mumbai quantum computer. The ansatz circuit (see Fig.~\ref{fig:4-site circuit}) for calculating the ground state of the 4-site Heisenberg Hamiltonian uses qubits $13$, $14$, $16$ and $19$ of the IBM Mumbai processor. These qubits are used in the ansatz implementation as the CNOT gate error for this set of qubits is low (of the order $10^{-3}$). Note that we have used $8192$ as the number of shots in our implementation. We have also applied the default readout error correction function while implementing our ansatz on the IBM processors \cite{Qiskit}.  
Fig.~\ref{fig:wo_correction} shows the result of our ansatz implementation for the 4-site Heisenberg Hamiltonian on IBM Mumbai. We observe from Fig.~\ref{fig:wo_correction} that the data points are noisy, and hence lead to an uneven $\frac{\partial{H}}{\partial{\phi}}$ Vs $\phi$ curve. Note that we use a five-point stencil to obtain the required derivative. The noise in the $\frac{\partial{H}}{\partial{\phi}}$ Vs $\phi$ data makes it difficult to extract the stationary points. So, we apply a Savitzky-Golay filter to our raw data, and then obtain the $\frac{\partial{H}}{\partial{\phi}}$ Vs $\phi$ plot (see Fig.~\ref{fig:with_correction}). We observe from Fig.~\ref{fig:with_correction} that the stationary points are at approximately $\phi=0$, $1.57$ and $3.14$, as expected.

\begin{center}
\begin{figure}[htbp!]
				\centerline{\includegraphics[width=0.35\textwidth,height=\textheight,keepaspectratio]{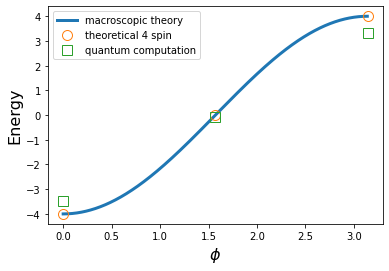}}
				\caption{Comparison of theoretical dispersion relation with that computed with the quantum processor. The continuous curve is for an infinite chain. }
				\label{fig:dispersion}
			\end{figure}
\end{center}

The theoretical dispersion relation is given by \cite{stancil2009spin}
\begin{equation}
    \frac{E_n}{J/2}=-N+4(1-\cos \phi_n).
\end{equation}
This is compared with the results from the quantum processor in Fig. \ref{fig:dispersion}. 
We see that the magnitude of the expected energy of our 4-site Heisenberg implementation (see also Fig.~\ref{fig:4-site with correction}) is slightly lower than the theoretical values. We attribute this reduced energy to the various noise sources (damping, dephasing and depolarizing) affecting the performance of IBM quantum processors. 

The number of CNOTs in our ansatz for the $4$-site Heisenberg Hamiltonian depends on the gate decomposition of the two unitaries $U_{\alpha}(\phi)$ and $U_\beta(\phi)$ (see Fig.~\ref{fig:4-site circuit}). Any two qubit unitary can be implemented using at most three CNOT gates \cite{vatan2004optimal}. Hence, the 4-site ansatz circuit can be implemented using $7$ CNOT gates. For comparison, the correlation function-based method requires $12$ CNOTs for implementing the time evolution of the 4-site Heisenberg Hamiltonian on a fully connected quantum processor. It further requires $2$ additional CNOTs to perform the hadamard test \cite{francis2020quantum}. Hence, our method reduces the number of CNOTs by $50\%$ in obtaining the single-magnon excited state of the 4-site Heisenberg spin chain. However, our Schmidt decomposition-based ansatz construction scheme does not scale efficiently. The number of CNOTs needed to create an arbitrary $n$ qubit state approximately scales as $2^n$. Hence, our ansatz implementation would require at most $256$ CNOT gates for obtaining the single-magnon excited states of an $8$-site Heisenberg Hamiltonian. Note that this estimate is the theoretical upper bound assuming full connectivity, and may increase in practical implementations due to sparse connectivity. On the other hand, the number of CNOT gates needed to implement the correlation function-based method scales linearly with $n$.

\begin{figure}[h!]
\centering
\begin{subfigure}{0.2\textwidth}
  \includegraphics[width=\linewidth]{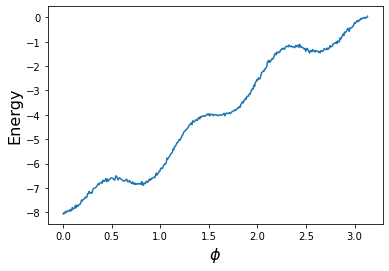}
  \caption{}
  \label{fig:no error}
\end{subfigure}\hfil
\begin{subfigure}{0.2\textwidth}
  \includegraphics[width=\linewidth]{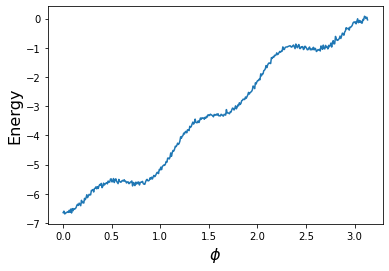}
  \caption{}
  \label{fig:0.001}
\end{subfigure}\hfil 
\begin{subfigure}{0.2\textwidth}
  \includegraphics[width=\linewidth]{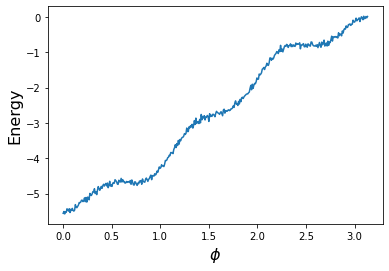}
  \caption{}
  \label{fig:0.0002}
\end{subfigure}\hfil 
\begin{subfigure}{0.2\textwidth}
  \includegraphics[width=\linewidth]{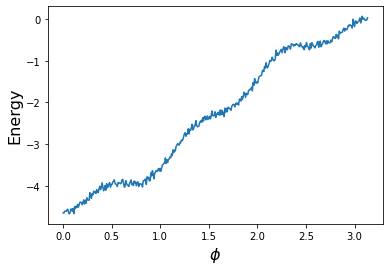}
  \caption{}
  \label{fig:0.0003}
\end{subfigure}\hfil 
\begin{subfigure}{0.2\textwidth}
  \includegraphics[width=\linewidth]{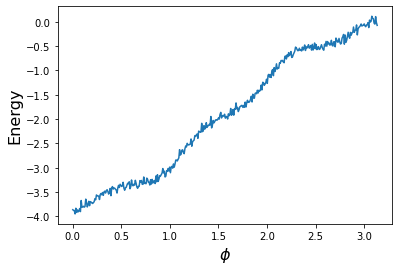}
  \caption{}
  \label{fig:0.004}
\end{subfigure}\hfil 
\caption{Energy Vs $\phi$ for different depolarizing error rates. a) ideal (no noise) implementation b) depolarizing error rate is $0.1\%$ c) depolarizing error rate is $0.2\%$ d) depolarizing error rate is $0.3\%$ e) depolarizing error rate is $0.4\%$. }
\label{fig:8-site_real}
\end{figure}

The presence of a large number of CNOTs in the 8-site ansatz makes our implementation infeasible on current NISQ devices. Hence, we implement our ansatz on a noiseless QASM simulator. Fig.~\ref{fig:no error} shows the energy vs $\phi$ plot for the 8-site Heisenberg Hamiltonian in the absence of any noise. 
We observe from Fig.~\ref{fig:no error} that the Energy vs $\phi$ curve is not smooth even in the absence of any noise. The non-smooth nature of the Energy vs $\phi$ curve can be attributed to the finite number of shots used in the QASM simulator for calculating the expected energy. As we increase $\phi$ in steps of $\frac{\pi}{800}$, the standard deviation is comparable to the energy difference between two neighboring values of $\phi$, thereby leading to a ``noisy'' Energy vs $\phi$ curve.

\begin{figure}[h!]
\centering
\begin{subfigure}{0.2\textwidth}
  \includegraphics[width=\linewidth]{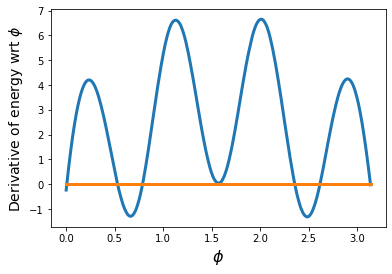}
  \caption{}
  \label{fig:no error_der}
\end{subfigure}\hfil
\begin{subfigure}{0.2\textwidth}
  \includegraphics[width=\linewidth]{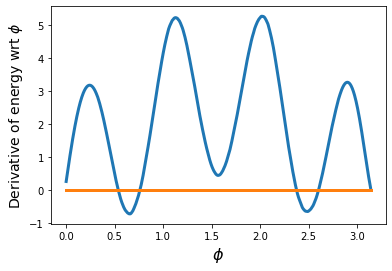}
  \caption{}
  \label{fig:0.001_der}
\end{subfigure}\hfil 
\begin{subfigure}{0.2\textwidth}
  \includegraphics[width=\linewidth]{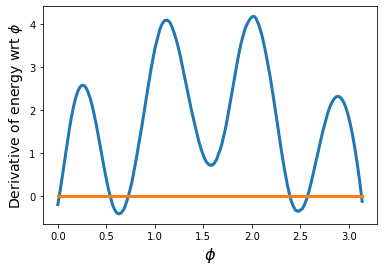}
  \caption{}
  \label{fig:0.0002_der}
\end{subfigure}\hfil 
\begin{subfigure}{0.2\textwidth}
  \includegraphics[width=\linewidth]{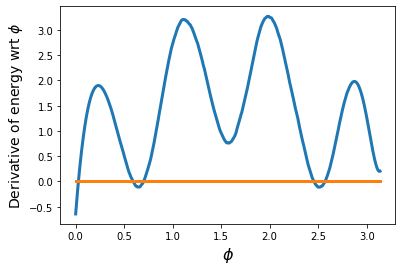}
  \caption{}
  \label{fig:0.0003_der}
\end{subfigure}\hfil 
\begin{subfigure}{0.2\textwidth}
  \includegraphics[width=\linewidth]{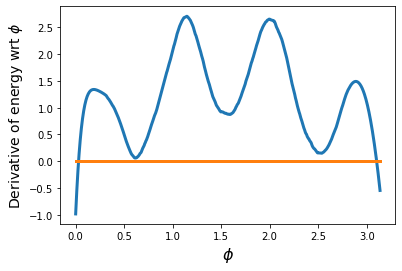}
  \caption{}
  \label{fig:0.004_der}
\end{subfigure}\hfil 
\caption{Derivative of energy Vs $\phi$ for different depolarizing error rates (filtered twice). a) ideal (no noise) implementation b) depolarizing error rate is $0.1\%$ c) depolarizing error rate is $0.2\%$ d) depolarizing error rate is $0.3\%$ e) depolarizing error rate is $0.4\%$}
\label{fig:with_noise}
\end{figure}

Next, we estimate the upper bound on the CNOT error rate that would give results for the ansatz of the 8-site Heisenberg Hamiltonian. For this purpose, we consider the case where the error in the CNOT gates of the IBM processors can be modeled as a depolarizing noise. Hence, we construct our 8-site ansatz on the QASM simulator \cite{Qiskit}, and deliberately add depolarizing error to the CNOT gates of our circuit. Fig.~\ref{fig:8-site_real} shows the performance of the 8-site ansatz for different depolarizing error rates. We again apply a Savitzky-Golay filter to our raw data, and then obtain the $\frac{\partial{H}}{\partial{\phi}}$ Vs $\phi$ plot (see Fig.~\ref{fig:with_noise}). 

\begin{center}
\begin{figure}[htbp!]
				\centerline{\includegraphics[width=0.35\textwidth,height=\textheight,keepaspectratio]{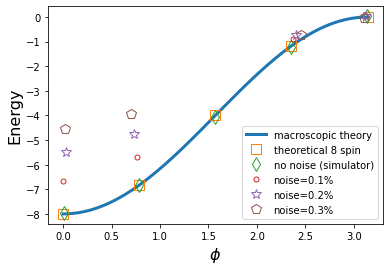}}
				\caption{Comparison of theoretical dispersion relation for an 8 spin chain with that computed with the quantum simulator as noise is added. }
				\label{fig:dispersion8}
			\end{figure}
\end{center}

The resulting estimated dispersion curve as noise is added is shown in Fig. \ref{fig:dispersion8}. It is clear that even a small amount of noise rapidly introduces significant errors in the dispersion curve.

Such low CNOT errors are not possible with the current technology. For example, ion-trap quantum computers designed by Honeywell offer an average CNOT fidelity of $99.5\%$.

An alternate approach to remove the requirement of such a precise CNOT is to reduce the number of CNOT gates. This CNOT number reduction can be done using an efficient quantum state preparation (QSP) algorithm. These QSP algorithms use the number of qubits, number of CNOTs, and depth of the circuit as a metric of efficiency. There are QSP algorithms that reduce the number of CNOTs in the circuit, but at the cost of increased depth and qubits \cite{zhang2021low, sun2021asymptotically}. Another insight towards efficient implementation of our ansatz is the fact that the coefficient matrix $C$ for the 8-site Heisenberg Hamiltonian is sparse (see Eq.~\eqref{eq: Coeff_8-site}. We can use a QRAM-based algorithm to construct the required circuit in which the number of CNOTs scales linearly \cite{de2021double}, provided the coefficient matrix is sparse. For example, the ansatz circuit for the 8-site Heisenberg can be implemented using only 32 CNOT gates. Thus, our work raises an interesting question about the efficiency of various quantum state preparation algorithms on a NISQ device, especially on a quantum processor with sparse connectivity.
\section{Conclusion}
We have presented an approach for calculating the single-magnon excited states of a Heisenberg spin chain that can be efficiently implemented on a quantum computer for small chains. The approach uses a wavefunction ansatz that is based on the intuitive properties of a propagating wave collective excitation. In particular, the ansatz is constructed from a superposition of single-flipped-spin states, with a linear progressive phase shift between states with adjacent flipped spins. Since the sought-after solution is a uniform collective excitation, the flipped-spin is equally likely to be on any site, and magnitudes of the amplitudes of all the single-flipped spin states are taken to be equal. Thus the incremental phase shift between states with adjacent flipped spins is the single parameter in the ansatz. We then proved that with this ansatz, the correct single-magnon excited states of the Heisenberg Hamiltonian correspond to stationary points of the energy vs phase shift curve.

To implement the method on a quantum processor, we implemented the ansatz for 4 and 8 spin chains using the quantum state preparation method described by Plesch and Brukner \cite{2011}. The quantum processor correctly identified the eigenstates for the 4 spin chain, but did not produce correct results for the 8 spin chain owing to noise and gate errors. We showed that this could be explained by considering how the number of CNOTs scales with the size of the spin chain for the quantum state preparation method used. 
Our work has shown the importance of designing efficient quantum state preparation algorithms suited for sparsely connected quantum processors.
\section*{Acknowledgements}
We acknowledge the use of IBM Quantum via the IBM Quantum Hub at NC State for this work. The views expressed are those of the authors and do not reflect the official policy or position of the IBM Quantum Hub at NC State, IBM or the IBM Quantum team. Shashank Ranu acknowledges financial support from the J. William Fulbright Foreign Scholarship Board and the Fulbright Commission in India (USIEF) through Fulbright Nehru Doctoral Research Fellowship 2021-2022. Shashank Ranu is thankful to Anil Prabhakar and Prabha Mandayam for helpful discussions. Both the authors thank Patrick Dreher and Alexander Kemper for their insightful comments.
\bibliography{ref}
\begin{widetext}
\appendix
\section{Ansatz for constructing the single-magnon state trial wavefunction for $8$-site Heisenberg Hamiltonian}\label{appendix:ansatz 8-site}
The coefficient matrix $C$ of the single-magnon state trial wavefunction of the 8-site spin chain  (see Eq.~\eqref{eq:excited_8-site}) is
\begin{equation}
    C_{\text{8-site}}=\frac{1}{2\sqrt{2}}\begin{pmatrix}
0 & 1 & e^{i\phi} & 0 & e^{i2\phi} & 0 & 0 & 0 & e^{i3\phi} & 0 & 0 & 0 & 0 & 0 & 0 & 0\\
e^{i4\phi} & 0 & 0 & 0 & 0 & 0 & 0 & 0 & 0 & 0 & 0 & 0 & 0 & 0 & 0 & 0\\
e^{i5\phi} & 0 & 0 & 0 & 0 & 0 & 0 & 0 & 0 & 0 & 0 & 0 & 0 & 0 & 0 & 0\\
0 & 0 & 0 & 0 & 0 & 0 & 0 & 0 & 0 & 0 & 0 & 0 & 0 & 0 & 0 & 0\\
e^{i6\phi} & 0 & 0 & 0 & 0 & 0 & 0 & 0 & 0 & 0 & 0 & 0 & 0 & 0 & 0 & 0\\
0 & 0 & 0 & 0 & 0 & 0 & 0 & 0 & 0 & 0 & 0 & 0 & 0 & 0 & 0 & 0\\
0 & 0 & 0 & 0 & 0 & 0 & 0 & 0 & 0 & 0 & 0 & 0 & 0 & 0 & 0 & 0\\
0 & 0 & 0 & 0 & 0 & 0 & 0 & 0 & 0 & 0 & 0 & 0 & 0 & 0 & 0 & 0\\
e^{i7\phi} & 0 & 0 & 0 & 0 & 0 & 0 & 0 & 0 & 0 & 0 & 0 & 0 & 0 & 0 & 0\\
0 & 0 & 0 & 0 & 0 & 0 & 0 & 0 & 0 & 0 & 0 & 0 & 0 & 0 & 0 & 0\\
0 & 0 & 0 & 0 & 0 & 0 & 0 & 0 & 0 & 0 & 0 & 0 & 0 & 0 & 0 & 0\\
0 & 0 & 0 & 0 & 0 & 0 & 0 & 0 & 0 & 0 & 0 & 0 & 0 & 0 & 0 & 0\\
0 & 0 & 0 & 0 & 0 & 0 & 0 & 0 & 0 & 0 & 0 & 0 & 0 & 0 & 0 & 0\\
0 & 0 & 0 & 0 & 0 & 0 & 0 & 0 & 0 & 0 & 0 & 0 & 0 & 0 & 0 & 0\\
0 & 0 & 0 & 0 & 0 & 0 & 0 & 0 & 0 & 0 & 0 & 0 & 0 & 0 & 0 & 0\\
0 & 0 & 0 & 0 & 0 & 0 & 0 & 0 & 0 & 0 & 0 & 0 & 0 & 0 & 0 & 0
\end{pmatrix}\label{eq: Coeff_8-site}
\end{equation}
The SVD of $C_{\text{8-site}}$ gives
\begin{eqnarray}
&&U=\begin{pmatrix}
e^{i3\phi} & 0 & 0 & 0 & 0 & 0 & 0 & 0 & 0 & 0 & 0 & 0 & 0 & 0 & 0 & 0\\
0 & \frac{e^{i4\phi}}{\sqrt{2}} & 0 & 0 & 0 & 0 & 0 & 0 & 0 & \frac{-e^{-i3\phi}}{\sqrt{2}} & 0 & 0 & 0 & \frac{e^{-i2\phi}}{\sqrt{6}} & 0 & \frac{-e^{-i\phi}}{2\sqrt{3}}\\
0 & \frac{e^{i5\phi}}{2} & 0 & 0 & 0 & 0 & 0 & 0 & 0 & 0 & 0 & 0 & 0 & 0 & 0 & \frac{\sqrt{3}}{2}\\
0 & 0 & 0 & 0 & 0 & 0 & 0 & 0 & 0 & 0 & 0 & 0 & 0 & 0 & 1 & 0\\
0 & \frac{e^{i6\phi}}{2} & 0 & 0 & 0 & 0 & 0 & 0 & 0 & 0 & 0 & 0 & 0 & \sqrt{\frac{2}{3}} & 0 & \frac{-e^{i\phi}}{2\sqrt{3}}\\
0 & 0 & 0 & 0 & 0 & 0 & 0 & 0 & 0 & 0 & 0 & 0 & 1 & 0 & 0 & 0\\
0 & 0 & 0 & 0 & 0 & 0 & 0 & 0 & 0 & 0 & 0 & 1 & 0 & 0 & 0 & 0\\
0 & 0 & 0 & 0 & 0 & 0 & 0 & 0 & 0 & 0 & 1 & 0 & 0 & 0 & 0 & 0\\
0 & \frac{e^{i7\phi}}{2} & 0 & 0 & 0 & 0 & 0 & 0 & 0 & \frac{1}{\sqrt{2}}  & 0 & 0 & 0 & \frac{-e^{i\phi}}{\sqrt{6}} & 0 & \frac{-e^{i2\phi}}{2\sqrt{3}}\\
0 & 0 & 0 & 0 & 0 & 0 & 0 & 0 & 1 & 0 & 0 & 0 & 0 & 0 & 0 & 0\\
0 & 0 & 0 & 0 & 0 & 0 & 0 & 1 & 0 & 0 & 0 & 0 & 0 & 0 & 0 & 0\\
0 & 0 & 0 & 0 & 0 & 0 & 1 & 0 & 0 & 0 & 0 & 0 & 0 & 0 & 0 & 0\\
0 & 0 & 0 & 0 & 0 & 1 & 0 & 0 & 0 & 0 & 0 & 0 & 0 & 0 & 0 & 0\\
0 & 0 & 0 & 0 & 1 & 0 & 0 & 0 & 0 & 0 & 0 & 0 & 0 & 0 & 0 & 0\\
0 & 0 & 0 & 1 & 0 & 0 & 0 & 0 & 0 & 0 & 0 & 0 & 0 & 0 & 0 & 0\\
0 & 0 & 1 & 0 & 0 & 0 & 0 & 0 & 0 & 0 & 0 & 0 & 0 & 0 & 0 & 0
\end{pmatrix}\nonumber\\\label{eq:U_8-site}\nonumber \\
&&\Sigma=\begin{pmatrix}
\frac{1}{\sqrt{2}} & 0 & 0 & 0 & 0 & 0 & 0 & 0 & 0 & 0 & 0 & 0 & 0 & 0 & 0 & 0\\
0 & \frac{1}{\sqrt{2}}  & 0 & 0 & 0 & 0 & 0 & 0 & 0 & 0 & 0 & 0 & 0 & 0 & 0 & 0\\
0 & 0 & 0 & 0 & 0 & 0 & 0 & 0 & 0 & 0 & 0 & 0 & 0 & 0 & 0 & 0\\
0 & 0 & 0 & 0 & 0 & 0 & 0 & 0 & 0 & 0 & 0 & 0 & 0 & 0 & 0 & 0\\
0 & 0 & 0 & 0 & 0 & 0 & 0 & 0 & 0 & 0 & 0 & 0 & 0 & 0 & 0 & 0\\
0 & 0 & 0 & 0 & 0 & 0 & 0 & 0 & 0 & 0 & 0 & 0 & 0 & 0 & 0 & 0\\
0 & 0 & 0 & 0 & 0 & 0 & 0 & 0 & 0 & 0 & 0 & 0 & 0 & 0 & 0 & 0\\
0 & 0 & 0 & 0 & 0 & 0 & 0 & 0 & 0 & 0 & 0 & 0 & 0 & 0 & 0 & 0\\
0 & 0 & 0 & 0 & 0 & 0 & 0 & 0 & 0 & 0 & 0 & 0 & 0 & 0 & 0 & 0\\
0 & 0 & 0 & 0 & 0 & 0 & 0 & 0 & 0 & 0 & 0 & 0 & 0 & 0 & 0 & 0\\
0 & 0 & 0 & 0 & 0 & 0 & 0 & 0 & 0 & 0 & 0 & 0 & 0 & 0 & 0 & 0\\
0 & 0 & 0 & 0 & 0 & 0 & 0 & 0 & 0 & 0 & 0 & 0 & 0 & 0 & 0 & 0\\
0 & 0 & 0 & 0 & 0 & 0 & 0 & 0 & 0 & 0 & 0 & 0 & 0 & 0 & 0 & 0\\
0 & 0 & 0 & 0 & 0 & 0 & 0 & 0 & 0 & 0 & 0 & 0 & 0 & 0 & 0 & 0\\
0 & 0 & 0 & 0 & 0 & 0 & 0 & 0 & 0 & 0 & 0 & 0 & 0 & 0 & 0 & 0\\
0 & 0 & 0 & 0 & 0 & 0 & 0 & 0 & 0 & 0 & 0 & 0 & 0 & 0 & 0 & 0
\end{pmatrix}\label{eq:sigma_8-site}
\end{eqnarray}
\begin{equation}
V=\begin{pmatrix}
0 & 1 & 0 & 0 & 0 & 0 & 0 & 0 & 0 & 0 & 0 & 0 & 0 & 0 & 0 & 0\\
\frac{e^{i3\phi}}{2} & 0 & 0 & 0 & 0 & 0 & 0 & 0 & 0 & \frac{-e^{i3\phi}}{\sqrt{2}} & 0 & 0 & 0 & \frac{-e^{i2\phi}}{\sqrt{6}} & 0 & \frac{-e^{i\phi}}{2\sqrt{3}}\\
\frac{e^{i2\phi}}{2} & 0 & 0 & 0 & 0 & 0 & 0 & 0 & 0 & 0 & 0 & 0 & 0 & 0 & 0 & \frac{\sqrt{3}}{2}\\
0 & 0 & 0 & 0 & 0 & 0 & 0 & 0 & 0 & 0 & 0 & 0 & 0 & 0 & 1 & 0\\
\frac{e^{i\phi}}{2} & 0 & 0 & 0 & 0 & 0 & 0 & 0 & 0 & 0 & 0 & 0 & 0 & \sqrt{\frac{2}{3}} & 0 & \frac{-e^{-i\phi}}{2\sqrt{3}}\\
0 & 0 & 0 & 0 & 0 & 0 & 0 & 0 & 0 & 0 & 0 & 0 & 1 & 0 & 0 & 0\\
0 & 0 & 0 & 0 & 0 & 0 & 0 & 0 & 0 & 0 & 0 & 1 & 0 & 0 & 0 & 0\\
0 & 0 & 0 & 0 & 0 & 0 & 0 & 0 & 0 & 0 & 1 & 0 & 0 & 0 & 0 & 0\\
\frac{1}{2} & 0 & 0 & 0 & 0 & 0 & 0 & 0 & 0 & \frac{1}{\sqrt{2}}  & 0 & 0 & 0 & \frac{-e^{-i\phi}}{\sqrt{6}} & 0 & \frac{-e^{-i2\phi}}{2\sqrt{3}}\\
0 & 0 & 0 & 0 & 0 & 0 & 0 & 0 & 1 & 0 & 0 & 0 & 0 & 0 & 0 & 0\\
0 & 0 & 0 & 0 & 0 & 0 & 0 & 1 & 0 & 0 & 0 & 0 & 0 & 0 & 0 & 0\\
0 & 0 & 0 & 0 & 0 & 0 & 1 & 0 & 0 & 0 & 0 & 0 & 0 & 0 & 0 & 0\\
0 & 0 & 0 & 0 & 0 & 1 & 0 & 0 & 0 & 0 & 0 & 0 & 0 & 0 & 0 & 0\\
0 & 0 & 0 & 0 & 1 & 0 & 0 & 0 & 0 & 0 & 0 & 0 & 0 & 0 & 0 & 0\\
0 & 0 & 0 & 1 & 0 & 0 & 0 & 0 & 0 & 0 & 0 & 0 & 0 & 0 & 0 & 0\\
0 & 0 & 1 & 0 & 0 & 0 & 0 & 0 & 0 & 0 & 0 & 0 & 0 & 0 & 0 & 0
\end{pmatrix}\label{eq:sigma_8-site}
\end{equation}
As a first step towards generating $\ket{\psi}_{\text{8-site}}$,  we apply a Hadamard gate to qubit $4$ to obtain the mapping shown in Eq.~\eqref{eq:4-site step 1a}, i.e
\begin{equation}
    \ket{00000000}\xrightarrow{I\otimes I\otimes I\otimes H\otimes I\otimes I \otimes I \otimes I} \frac{1}{\sqrt{2}}\left(\ket{0000}+\ket{0001}\right)\otimes\ket{0000}.\label{eq: 8-site step 1}
\end{equation}
A single CNOT gate between qubit $4$ and $8$ implements the required transformation of Eq.~\eqref{eq:4-site step 2a} in the case of $\ket{\psi}_{\text{8-site}}$.
\begin{equation}
\frac{1}{\sqrt{2}}\left(\ket{0000}+\ket{0001}\right)\otimes\ket{0000}\xrightarrow{CNOT_{2\to 4}}\frac{1}{\sqrt{2}}(\ket{0000}\otimes\ket{0000}+\ket{0001}\otimes\ket{0001}),\label{eq:8-site step 2}
 \end{equation}
where $CNOT_{4\to 8}$ represents a CNOT gate with qubit $4$ as control qubit and qubit $8$ as target qubit. Next, we find that the unitary matrix ($U_\alpha(\phi)$) that does the transformation of Eq.~\eqref{eq:4-site step 3a} when $\ket{\psi}_{\text{8-site}}$ is the target state is identical to $U$ (see Eq.~\eqref{eq:U_8-site}). $U_\alpha(\phi)$ acts on the state shown in Eq.~\eqref{eq:8-site step 2} as shown below
\begin{eqnarray}
 \frac{1}{\sqrt{2}}(\ket{0000}\otimes\ket{0000}+\ket{0001}\otimes\ket{0001})\xrightarrow{U_\alpha(\phi)}\frac{e^{i3\phi}}{\sqrt{2}}&&\ket{0000}\otimes\ket{0000}+\frac{1}{2\sqrt{2}}\big(e^{i4\phi}\ket{0001}\otimes\ket{0001}\nonumber\\
 &&+e^{i5\phi}\ket{0010}\otimes\ket{0001}+e^{i6\phi}\ket{0100}\otimes\ket{0001}\nonumber \\
 &&+e^{i7\phi}\ket{1000}\otimes\ket{0001}\big)
 \label{eq:8-site step 3}
\end{eqnarray}
Lastly, the unitary matrix $U_{\beta}(\phi)$ must be equal to $V^{\dagger}$ (see Eq.~\eqref{eq:sigma_8-site}) that carries out the transformation of Eq.~\eqref{eq:4-site step 4a} when $\ket{\psi}_{\text{8-site}}$ is the target state. Action of $U_\beta(\phi)$ on the state shown in Eq.~\eqref{eq:8-site step 3} gives $\ket{\psi}_{\text{8-site}}$.
\end{widetext}
\end{document}